%% file: main.tex
\documentclass[conference]{IEEEtran}
\IEEEoverridecommandlockouts
\usepackage{cite}
\usepackage{amsmath,amssymb,amsfonts}
\usepackage{graphicx}
\usepackage{booktabs}
\usepackage{siunitx}
\usepackage{xcolor}
\usepackage{tikz}
\usetikzlibrary{arrows.meta,patterns}
\usepackage[hidelinks]{hyperref}
\usepackage{colortbl}
\definecolor{headblue}{RGB}{221,235,247}
\input{numbers2.tex}

\definecolor{groupgray}{RGB}{242,242,242}
\definecolor{bettergreen}{RGB}{226,240,217}
\usepackage{multirow}
\begin{document}
\bstctlcite{IEEEexample:BSTcontrol}

\title{mmWave UE Positioning Under Blockage:\\
Limits for ISAC-Assisted Handover}

\author{
\IEEEauthorblockN{Seyed Bagher Hashemi Natanzi\IEEEauthorrefmark{1}\thanks{*Corresponding author: snatanzi@wpi.edu}, Bo Tang\IEEEauthorrefmark{2}}
\IEEEauthorblockA{\IEEEauthorrefmark{1}\IEEEauthorrefmark{2}Department of Electrical and Computer Engineering, Worcester Polytechnic Institute (WPI), USA}
\IEEEauthorblockA{Email: \{snatanzi, btang1\}@wpi.edu}
}

\maketitle
    \begin{tikzpicture}[remember picture, overlay]
        \node[anchor=north, yshift=-0.5cm] at (current page.north) {\fbox{\parbox{\textwidth}{\centering\small {\color{red}This work has been submitted to the IEEE Conference for possible publication. Copyright may be transferred without notice, after which this version may no longer be accessible.}}}};
    \end{tikzpicture}
\begin{abstract}

Proactive handover can save a millimeter-wave (mmWave) link before a
passing body blocks it, but only if the network knows where the user is.
We ask whether the radio units that serve a user can localize it
accurately enough from its uplink sounding reference signals. We derive
position error bounds that include synchronization and array calibration
errors, with and without reflected paths known from a digital twin, and
test two estimators in a ray-traced street canyon on held-out scenarios.
Three findings stand out. First, with line of sight to both radio units,
array calibration, not bandwidth, limits the bound. Second, blockage
degrades the line-of-sight-only bound sharply, whereas known reflections
keep the map-aided bound low. Third, a simple dominant-path estimator
fails where it matters most: its errors grow sharply around blockage,
and proactive handover becomes worse than the 3GPP baseline. A map-aided
multipath tracker cuts the median error from \pEstMedFourHundred{} to
\pTrkMed\,cm and removes this significant penalty at link margins of
\SI{15}{dB} and above, but does not yet beat the baseline.
\end{abstract}


\section{Introduction}
Millimeter-wave (mmWave) bands provide the bandwidth that 5G and 6G
networks need for high data rates, but their links are fragile: a
pedestrian or vehicle crossing the line of sight (LoS) attenuates the
signal by tens of decibels within milliseconds
\cite{rappaport2013mmwave,maccartney2017blockage}. Reactive handover
cannot avoid such outages. Integrated sensing and communication (ISAC) and
other sensing-aided approaches therefore aim to predict blockage and
hand the user over in advance
\cite{liu2022isac,charan2021vision,li2025lookbefore}.

Predicting a blockage means predicting whether a body will cross the
line between a radio unit and the user. The prediction can thus be no
better than the network's knowledge of where the user is. In a companion
study \cite{natanzi2027companion}, we found that a blockage-aware
handover planner outperforms the 3GPP A5 event, a reactive trigger that
hands over when the serving cell falls below one threshold and a
neighbor exceeds another, only if
the user position is known with an error of about \pBreakEven\,cm (per
horizontal axis), and that a meter-level error removes the benefit
entirely.

Can the network meet this requirement with the infrastructure it
already has? 5G NR supports uplink positioning from sounding reference
signals (SRS) \cite{3gpp38855}, and recent street-scale studies with
several base stations report errors below a meter
\cite{dong2026deployable}. Position error bounds for mmWave systems
suggest that much better accuracy is possible in principle
\cite{shahmansoori2018position,abushaban2018bounds}, but they typically
consider ideal hardware or a single impairment in static scenes
\cite{keskin2022optimal}. It remains unclear which effect
limits decimeter accuracy in practice, how blockage changes it, how
close practical estimators come, and whether their errors are small
enough when handover decisions are made.

This paper addresses these questions in an urban street canyon, a street
lined with buildings on both sides, served by two O-RAN radio units
(O-RUs). Our contributions are:
\begin{itemize}
\item A positioning analysis for blockage-aware handover: a
single-epoch equivalent Fisher information analysis with
synchronization and array calibration errors as Bayesian priors, two
models of a blocked LoS, and map-aided information from reflected paths
with known geometry.
\item Two practical uplink estimators: a dominant-path estimator with an
extended Kalman filter (EKF), whose single-epoch front end is compared
with the bound by decomposing its error into bias and random spread, and
a map-aided multipath tracker that associates extracted paths with
virtual anchors of the digital twin.
\item An analysis of positioning errors around blockage onsets and a
closing experiment that feeds estimated and synthetic bound-level
positions into a blockage-aware handover planner. All code and the scripts that generate every number are open source.\footnote{\url{https://github.com/CLIS-WPI/sense2connect}}
\end{itemize}

\section{Related Work}
\textit{Bounds for mmWave positioning.} Fisher information analyses
show that the angle and delay resolution of mmWave arrays support
sub-decimeter accuracy \cite{shahmansoori2018position,abushaban2018bounds},
building on the equivalent Fisher information framework for wideband
localization \cite{shen2010fundamental}. Multipath components with known
geometry carry position information as virtual anchors
\cite{leitinger2015evaluation,witrisal2016high}, and multipath tracking
associates them with such anchors probabilistically
\cite{meyer2018message}. We add hardware errors as priors and evaluate
bounds and estimators under traffic-driven blockage.

\textit{Uplink SRS positioning.} 3GPP specifies uplink time- and
angle-based positioning \cite{3gpp38855}. Dong \emph{et al.}
\cite{dong2026deployable} combine LoS/non-LoS gating, dominant-path angle
and delay estimation, and multi-base-station EKF fusion, reach sub-meter
errors in a ray-traced campus, and point to reflections as virtual
anchors for future work.

\textit{Sensing-aided handover.} Vision-, radar-, and ISAC-aided
handover schemes predict blockage or user motion
\cite{charan2021vision,li2025lookbefore}; our companion study
\cite{natanzi2027companion} identified the user position as a binding
requirement.

\section{System Model and Problem Formulation}
\subsection{Signal Model}
A single-antenna user equipment (UE) at horizontal position $\mathbf p$
at known height (distances and directions are evaluated at the 3D point
$[\mathbf p^{\mathsf T},z_{\mathrm{UE}}]^{\mathsf T}$) transmits an uplink SRS that is received by a set $\mathcal C$ of O-RUs,
each with an $M$-element array. At O-RU $c$, subcarrier frequency
$f_n$, and element $m$, the received sample is
\begin{equation}
y_{c,n,m}=\sum_{\ell}\beta_{c,\ell}\,
e^{-j2\pi f_n(\tau_{c,\ell}+\delta_c+b)}\,
e^{j\psi_{c,m}}\,s_m(\mathbf u_{c,\ell})+w_{c,n,m},
\label{eq:signal}
\end{equation}
where path $\ell$ has complex gain $\beta_{c,\ell}$ (including blockage
loss), delay $\tau_{c,\ell}$, and arrival direction
$\mathbf u_{c,\ell}$ with steering response $s_m$; $\delta_c$ is the time
offset of O-RU $c$, $b$ the UE clock bias, $\psi_{c,m}$ the calibration
phase error of element $m$, and $w_{c,n,m}\sim\mathcal{CN}(0,\sigma_w^2)$
independent thermal noise. Hardware
errors are random with $\delta_c\sim\mathcal N(0,\sigma_{\mathrm{sync}}^2)$
and $\psi_{c,m}\sim\mathcal N(0,\sigma_\phi^2)$.

\subsection{Geometry and Blockage}
The LoS delay and angles are functions of $\mathbf p$ and the O-RU
position $\mathbf p_c$, e.g., $\tau_{c,0}=\|\mathbf p-\mathbf p_c\|/c_0$.
A specular reflection on a known surface is an LoS path from the virtual
anchor $\tilde{\mathbf p}_{c,\ell}$, the image of $\mathbf p_c$ in that
surface, so that $\tau_{c,\ell}=\|\mathbf p-\tilde{\mathbf p}_{c,\ell}\|/c_0$
\cite{leitinger2015evaluation}. Moving bodies attenuate every path
segment according to a blockage model; we use 3GPP model~B, which
treats each body as a screen with knife-edge diffraction at its edges
\cite{3gpp38901}. An O-RU is \emph{blocked} for a UE when the loss of its LoS path exceeds
a threshold. Model~B reduces the power of a blocked LoS
but keeps its geometry, whereas the field that actually arrives
diffracts around the blocker. We therefore consider two models of a
blocked LoS: \emph{kept}, in which its delay and angles still follow
$\mathbf p$ (optimistic), and \emph{biased}, in which they carry an
unknown bias. In the biased model, a bound treats them as free parameters, and an
estimator observes a blocked LoS whose delay and direction deviate from
the geometry of $\mathbf p$.

\subsection{Positioning Requirement of Blockage-Aware Handover}
A blockage-aware handover policy $\pi$ predicts the SNR of each cell
from the tracked blockers and the reported UE position
$\hat{\mathbf p}_t=\mathbf p_t+\mathbf e_t$, and selects the serving cell
accordingly. Its outage time $J(\pi;\{\hat{\mathbf p}_t\})$ counts the
time below the SNR required for the service rate, including the
interruptions of its handovers. Since the policy aims to outperform a
reactive reference policy $\pi_0$ (here, the 3GPP A5 event), the
positioning requirement is that the error process $\{\mathbf e_t\}$ keeps
\begin{equation}
J\bigl(\pi;\{\hat{\mathbf p}_t\}\bigr)\le J(\pi_0).
\label{eq:req}
\end{equation}
This requirement depends on the error at the times the policy acts, not
only on its overall statistics. We therefore examine (i) what limits
the position error bound, (ii) how blockage and map knowledge change it,
(iii) how a practical estimator compares, and (iv) whether its errors
satisfy \eqref{eq:req}.

\section{Position Error Bound and Estimator}
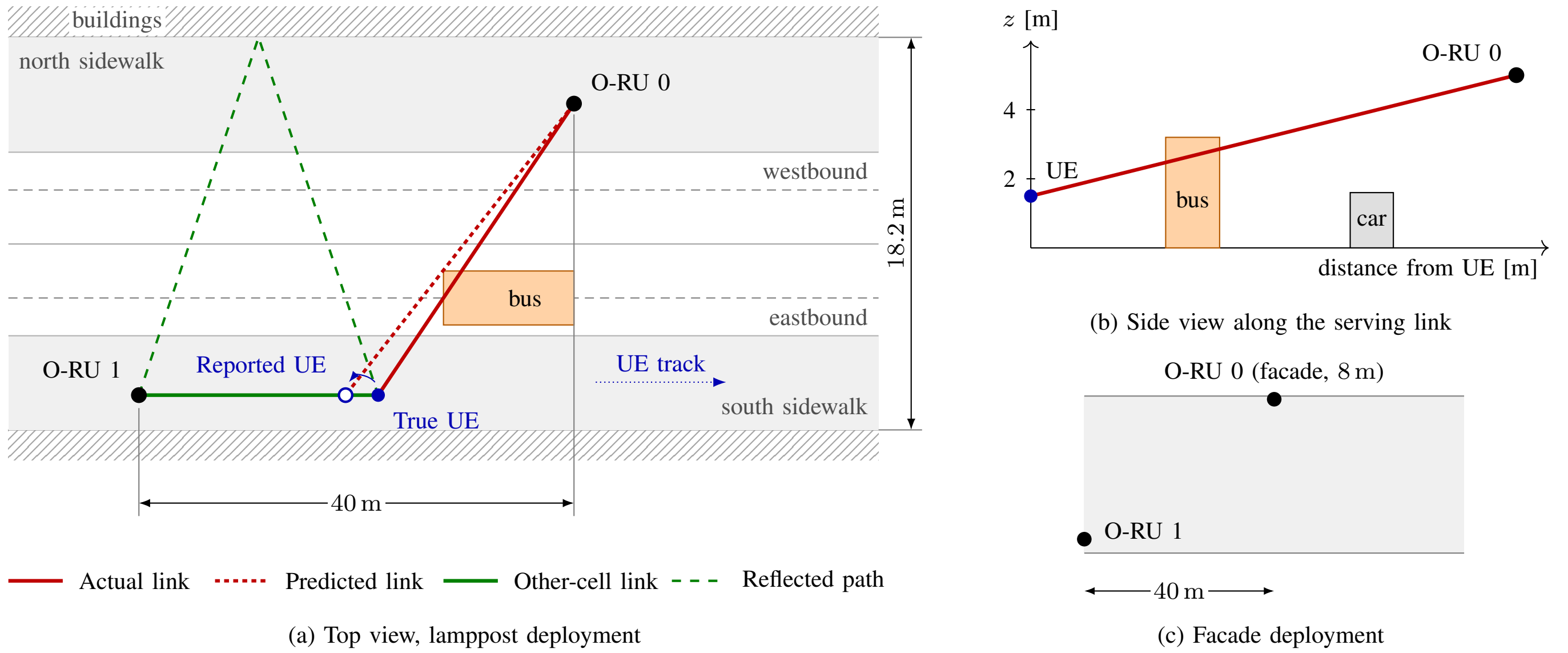
\begin{figure*}[t]
\centering
\begin{minipage}[b]{0.60\textwidth}
\centering
\begin{tikzpicture}[x=1.26mm,y=2.5mm,font=\footnotesize]
  \fill[pattern=north east lines,pattern color=gray!70] (-40,9.57) rectangle (40,11.0);
  \fill[pattern=north east lines,pattern color=gray!70] (-40,-8.61) rectangle (40,-10.0);
  \draw[gray] (-40,9.57)--(40,9.57); \draw[gray] (-40,-8.61)--(40,-8.61);
  \node[gray!60!black,fill=white,inner sep=1pt] at (-30,10.3) {buildings};
  \fill[gray!12] (-40,4.25) rectangle (40,9.57);
  \fill[gray!12] (-40,-8.61) rectangle (40,-4.25);
  \draw[gray!60] (-40,4.25)--(40,4.25); \draw[gray!60] (-40,-4.25)--(40,-4.25);
  \draw[densely dashed,gray] (-40,2.5)--(40,2.5); \draw[densely dashed,gray] (-40,-2.5)--(40,-2.5);
  \draw[gray!70] (-40,0)--(40,0);
  \node[gray!60!black,anchor=south east] at (40,2.5) {westbound};
  \node[gray!60!black,anchor=north east] at (40,-2.5) {eastbound};
  \node[gray!60!black,anchor=north west] at (-40,9.4) {north sidewalk};
  \node[gray!60!black,anchor=south east] at (40,-8.4) {south sidewalk};
  \draw[fill=orange!35,draw=orange!70!black] (0,-3.75) rectangle (12,-1.25);
  \node at (7.5,-2.5) {bus};
  \draw[very thick,red!75!black] (12,6.5)--(-6,-7);
  \draw[red!75!black,line width=1.1pt,dash pattern=on 1.6pt off 1.4pt] (12,6.5)--(-9,-7);
  \draw[very thick,green!50!black] (-28,-7)--(-6,-7);
  \draw[thick,green!50!black,dashed] (-28,-7)--(-17,9.57)--(-6,-7);
  \draw[-{Latex[length=1.5mm]},blue!70!black,densely dotted] (14,-6.4)--(26,-6.4);
  \node[blue!70!black,anchor=south] at (20,-6.4) {UE track};
  \fill[blue!70!black] (-6,-7) circle (2.2pt);
  \draw[blue!70!black,thick,fill=white] (-9,-7) circle (2.2pt);
  \draw[-{Latex[length=1.3mm]},blue!70!black] (-6.3,-6.4) to[bend right=50] (-8.7,-6.4);
  
  \node[blue!70!black,anchor=north west] at (-5.6,-7.3) {True UE};
  \node[blue!70!black,anchor=south east] at (-9.8,-6.7) {Reported UE};
  \fill (12,6.5) circle (2.6pt); \node[anchor=south west] at (12.6,6.6) {O-RU 0};
  \fill (-28,-7) circle (2.6pt); \node[anchor=south east] at (-28.6,-6.7) {O-RU 1};
  \draw[{Latex[length=1.3mm]}-{Latex[length=1.3mm]}] (-28,-12)--(12,-12);
  \draw[gray] (-28,-7.6)--(-28,-12.6); \draw[gray] (12,6)--(12,-12.6);
  \node[fill=white,inner sep=1pt] at (-8,-12) {\SI{40}{m}};
  \draw[{Latex[length=1.3mm]}-{Latex[length=1.3mm]}] (43,-8.61)--(43,9.57);
  \draw[gray] (40,-8.61)--(44,-8.61); \draw[gray] (40,9.57)--(44,9.57);
  \node[rotate=90,anchor=south] at (43.3,0.5) {\SI{18.2}{m}};
  \begin{scope}[shift={(-40,-15.6)}]
    \draw[very thick,red!75!black] (0,0)--(5,0); \node[anchor=west] at (5.5,0) {Actual link};
    \draw[red!75!black,line width=1.1pt,dash pattern=on 1.6pt off 1.4pt] (19,0)--(24,0); \node[anchor=west] at (24.5,0) {Predicted link};
    \draw[very thick,green!50!black] (40,0)--(45,0); \node[anchor=west] at (45.5,0) {Other-cell link};
    \draw[thick,green!50!black,dashed] (61,0)--(66,0); \node[anchor=west] at (66.5,0) {Reflected path};
  \end{scope}
\end{tikzpicture}\\[0.5mm]
{\footnotesize (a) Top view, lamppost deployment}
\end{minipage}\hfill
\begin{minipage}[b]{0.37\textwidth}
\centering
\begin{tikzpicture}[x=2.5mm,y=4mm,font=\footnotesize]
  \draw[->] (0,0)--(24,0) node[anchor=north east,align=right] {distance from UE [m]};
  \draw[->] (0,0)--(0,6) node[anchor=south] {$z$ [m]};
  \foreach \z in {2,4} {\draw (-0.2,\z)--(0.2,\z); \node[anchor=east] at (-0.2,\z) {\z};}
  \draw[fill=orange!35,draw=orange!70!black] (6.25,0) rectangle (8.75,3.2);
  \node at (7.5,1.4) {bus};
  \draw[fill=gray!25] (14.8,0) rectangle (16.8,1.6);
  \node at (15.8,0.8) {car};
  \draw[very thick,red!75!black] (0,1.5)--(22.5,5);
  \fill[blue!70!black] (0,1.5) circle (2.2pt); \node[anchor=south west] at (0.2,1.7) {UE};
  \fill (22.5,5) circle (2.6pt); \node[anchor=south east] at (22.3,5.1) {O-RU 0};
\end{tikzpicture}\\[0.5mm]
{\footnotesize (b) Side view along the serving link}\\[2mm]
\begin{tikzpicture}[x=0.55mm,y=1.0mm,font=\footnotesize]
  \fill[gray!12] (-40,-8.61) rectangle (40,9.57);
  \draw[gray] (-40,9.57)--(40,9.57); \draw[gray] (-40,-8.61)--(40,-8.61);
  \fill (0,9.2) circle (2.4pt); \node[anchor=south] at (0,9.8) {O-RU 0 (facade, \SI{8}{m})};
  \fill (-40,-7) circle (2.4pt); \node[anchor=west] at (-38,-6) {O-RU 1};
  \draw[{Latex[length=1.2mm]}-{Latex[length=1.2mm]}] (-40,-13)--(0,-13);
  \node[fill=white,inner sep=1pt] at (-20,-13) {\SI{40}{m}};
\end{tikzpicture}\\[0.5mm]
{\footnotesize (c) Facade deployment}
\end{minipage}
\caption{Scenario geometry (dimensions indicated; axes scaled independently). (a) Two neighboring cells share the
street; O-RU~0 and O-RU~1 are on \SI{5}{m} lampposts, \SI{40}{m} apart
along the street. The
true serving link (solid red) crosses the bus, whereas the link predicted
from the reported UE position (dotted) passes it, so a position error
turns into a missed blockage prediction; green: link to the other cell
and one of its facade reflections (dashed). (b) The LoS passes through a
bus but above a car. (c) In the facade deployment, O-RU~0 is mounted on
the north facade at \SI{8}{m}.}
\label{fig:scenario}
\end{figure*}

\subsection{Single-Epoch Position Error Bound}
Collect the channel parameters of all O-RUs at one epoch in
$\boldsymbol\eta$ (delays, angles, complex gains) and the hardware and
timing errors in $\boldsymbol\nu=[\{\delta_c\},b,\{\psi_{c,m}\}]$. The
Fisher information of $\boldsymbol\eta$ follows from \eqref{eq:signal}
by the Slepian--Bangs formula; we compute it exactly by factorizing its
frequency and antenna sums. With $\mathbf T=\partial\boldsymbol\eta/\partial\mathbf p$,
the equivalent Fisher information of the position is
\begin{equation}
\mathbf J_{\mathbf p}=\mathbf T^{\mathsf T}\mathbf J_{\boldsymbol\eta}\mathbf T
-\mathbf T^{\mathsf T}\mathbf J_{\boldsymbol\eta\boldsymbol\nu}
\bigl(\mathbf J_{\boldsymbol\nu\boldsymbol\nu}+\boldsymbol\Lambda\bigr)^{-1}
\mathbf J_{\boldsymbol\nu\boldsymbol\eta}\mathbf T,
\label{eq:efim}
\end{equation}
where all blocks are those remaining after the channel nuisance
parameters have been eliminated jointly by a Schur complement: the
complex gains of all paths; in the \emph{LoS-only} bound, the delays and
angles of all reflected paths; and in the biased model, the delay and
angles of a blocked LoS. The second term eliminates the timing and
hardware errors in $\boldsymbol\nu$. $\boldsymbol\Lambda$ holds the prior precisions of the
O-RU time offsets ($\sigma_{\mathrm{sync}}$) and of the element phases
($\sigma_\phi$); the UE clock bias has no prior (time difference of
arrival, TDoA). In the \emph{map-aided} bound, the delays and angles of the reflected
paths are not nuisance parameters: their geometry is known from the digital twin,
and their entries of $\mathbf T$ follow analytically from the virtual
anchors. Finite differences at \pDerivFdNumUe{} random positions, an independent NumPy implementation, and a brute-force Fisher matrix confirm the derivatives and \eqref{eq:efim} to relative errors of \pDerivFdMaxRel, \pFimJobErr, and \pFimToyErr. The position
error bound is $\mathrm{PEB}=\sqrt{\mathrm{tr}(\mathbf J_{\mathbf p}^{-1})}$.
Because the hardware errors enter through priors, it is a hybrid
Cram\'er--Rao bound: it bounds the root-mean-square (RMS) error of
unbiased single-epoch estimates for a given geometry, averaged over the
noise and the hardware priors; it is neither a median error nor a
guarantee that an estimator attains it, a biased estimator can have a
smaller spread, and it excludes the temporal information of
tracking.

\subsection{Dominant-Path Estimator}
Per O-RU and epoch, a zero-padded two-dimensional DFT over the array
selects the strongest beam; a Hann-windowed delay transform of the
beamformed signal gives the delay of the dominant path, refined by
parabolic interpolation and followed by an angle refinement at that
delay. A peak-to-second-peak test flags reflected dominant paths.
Because a planar array of isotropic elements cannot distinguish the two
sides of its plane, each angle yields two direction hypotheses; the
estimator keeps the joint hypothesis that is consistent across O-RUs,
lies in the walkable area of the digital-twin map, and passes the gate
of a constant-velocity EKF, which fuses weighted least-squares fixes
over time.

\subsection{Map-Aided Multipath Tracker}
Per O-RU and epoch, a sparse extraction with continuous refinement
returns up to eight multipath components with variances from the local Fisher information. The sources
are the LoS and every single- and double-bounce virtual anchor of the
street-facing building faces and the ground that is geometrically valid
at the predicted position. Loopy belief propagation computes the
probabilities that a source generated a component or that the component
is clutter, with constant visibility priors and a blocked mode that
models a detoured LoS by an inflated covariance \cite{meyer2018message};
a probabilistic-data-association EKF with a
constant-velocity model then updates the position, velocity, UE clock
bias, and O-RU time offsets, and projects the position onto the
walkable area. A grid search over the walkable area initializes the
track and re-initializes it after a loss of association.

\section{Experiments and Results}

\begin{table*}[t]
\centering
\caption{Single-epoch estimator error at fixed geometries
(\pValNumDraws{} draws each, main hardware, TDoA, no EKF) versus the
single-epoch LoS-only PEB (biased model for a blocked LoS); all values in
cm; RMSE$^2$ = bias$^2$ + centered RMS$^2$
(Eq.~\eqref{eq:decomp}).}
\label{tab:val}
\setlength{\tabcolsep}{4pt}
\setlength{\aboverulesep}{0pt}\setlength{\belowrulesep}{0pt}
\setlength{\extrarowheight}{1.5pt}
\begin{tabular}{llr|rrrr|rrrr}
\toprule
\rowcolor{headblue}
 & & & \multicolumn{4}{c|}{\pBwC\,MHz} & \multicolumn{4}{c}{\pBwA\,MHz}\\
\rowcolor{headblue}
Mount & LoS state & UE $x$ [m] & PEB & bias & centered & RMSE & PEB & bias & centered & RMSE\\
\midrule
lamppost & both LoS & -20.0 & 1.5 & 1.3 & 1.7 & 2.2 & 1.5 & 7.7 & 7.4 & 10.7 \\
lamppost & both LoS & 20.0 & 4.2 & 1496.1 & 55.4 & 1497.1 & 4.5 & 875.6 & 345.0 & 941.2 \\
lamppost & blocked (O-RU 0) & 21.9 & 46.3 & 301.3 & 4.9 & 301.3 & 91.5 & 286.1 & 9.3 & 286.3 \\
facade & both LoS & -20.0 & 3.3 & 31.9 & 4.8 & 32.2 & 3.4 & 30.0 & 16.1 & 34.1 \\
facade & both LoS & 20.0 & 6.7 & 88.6 & 6.1 & 88.8 & 7.1 & 94.0 & 9.4 & 94.5 \\
facade & blocked (O-RU 0) & 21.9 & 90.8 & 189.2 & 7.0 & 189.3 & 208.0 & 185.0 & 10.4 & 185.3 \\
\bottomrule
\end{tabular}
\end{table*}

\subsection{Setup}
\textit{Scenario.} We use the street canyon of
\cite{natanzi2027companion} (Fig.~\ref{fig:scenario}): buildings
\pStreetWidth\,m apart, two \pLaneWidth-m lanes, and sidewalks on both
sides; actors and UEs move on a periodic section of \pStreetPeriod\,m.
Two O-RUs with \pArray{} half-wavelength arrays at \pCarrier\,GHz form
two neighboring cells that share this section: O-RU~0 on the north side,
either on a lamppost (\pOruZeroLampZ\,m high) or on a facade
(\pOruZeroFacZ\,m), and O-RU~1 on a \pOruOneLampZ-m lamppost on the south
sidewalk, \pOruOffset\,m further along the street. Pedestrian UEs at
\pUeHeight\,m walk along the south sidewalk at \pUeSpeed\,m/s. Each
\pRunDuration-s run has low or high traffic: \pVehLaneKmLow{} or
\pVehLaneKmHigh{} vehicles per lane and km (cars, buses, trucks at
\pVehSpeed\,m/s) and \pPedWalkersLow{} or \pPedWalkersHigh{} pedestrians
on the sidewalks plus \pPedCrossingLow{} or \pPedCrossingHigh{} crossing. Paths are traced with Sionna RT
\cite{hoydis2023sionnart,sionna} (specular, order up to \pMaxDepth, at
most \pMaxPaths{} paths per link).

\textit{Service model.} The SNR of a cell adds, over all traced paths,
the path power with the full array gain after its model-B loss; this is an optimistic approximation, since no single beam collects all paths with full gain. The UE is in outage when
this total remaining power gives an SNR below
$\gamma_{\mathrm{req}}=\pSNRreq$\,dB, the value needed for
\SI{400}{Mbit/s}. As in \cite{natanzi2027companion}, the link margin is the median unblocked best-cell SNR above $\gamma_{\mathrm{req}}$ (3GPP budget: \pRefMargin\,dB, Ref.), swept from 0 to \SI{30}{dB} by a common extra path loss. During \SI{10}{dB} LoS events, the
strongest surviving path of the same cell lies a median
\pSameCellLoss\,dB below the unblocked LoS, so the margin decides
whether the same cell can still serve the UE.

\textit{Handover model.} The policy is the planner of
\cite{natanzi2027companion}: every \pEpoch\,s it receives the blocker
tracks from ISAC sensing and the reported UE position, predicts the SNR of each cell by
applying model~B on the line between each O-RU and the reported
position, and chooses the cell sequence that minimizes the predicted
outage over a horizon of \pPlanH\,s, with an interruption of \pTauHO\,ms
per handover; decisions take effect after a control-loop delay of
\pLoopDelay\,ms, and the sensing overhead of \pOverhead{} of the
resources is charged to it. The A5 thresholds and time-to-trigger are
tuned per margin on the tuning seeds; the tuned time-to-trigger,
\SI{40}{ms}, is the smallest value of the grid. With perfect blocker tracks and a white
Gaussian UE position error, the planner satisfies \eqref{eq:req} only up to a
per-axis standard deviation of about \pBreakEven\,cm, i.e., an RMS
distance error of about \pBreakEvenRms\,cm \cite{natanzi2027companion}.

\textit{Blocked LoS.} An O-RU is blocked for a UE at an epoch when the
model-B loss of its LoS path is at least \pBlockedDb\,dB. In the biased
model, the signal fed to the estimator replaces the blocked LoS by a
detour around its dominant blocker (the blocker with the largest model-B
loss): with $\mathbf q$ the shorter of the detour points over the top
and past either side of the blocker box, next to the point of the LoS
closest to the blocker, the path keeps the attenuated complex gain of
the LoS but has delay
$(\|\mathbf q-\mathbf p_c\|+\|\mathbf p-\mathbf q\|)/c_0$ and arrival
direction along $\mathbf q-\mathbf p_c$; reflected paths are only
attenuated.

\textit{Signals and hardware.} The UE transmits one SRS symbol per
\pEpoch-s epoch at \pTxPower\,dBm, spread uniformly over the active
subcarriers; the O-RU noise figure is
\pNoiseFig\,dB. Numerology~\pNumerology{} (\pScs\,kHz) with \pBwA,
\pBwB, and \pBwC\,MHz carriers uses \pNscA, \pNscB, and \pNscC{} active
subcarriers \cite{3gpp38101_2}. We sweep
$\sigma_{\mathrm{sync}}$ of \pSyncListShown\,ns (drawn once per run, a
bias that the EKF cannot average out; 40 realizations over the held-out
runs) and $\sigma_\phi$ of \pPhiList\,deg
(fixed per O-RU and run). The main configuration uses TDoA, \pBwC\,MHz,
$\sigma_{\mathrm{sync}}=\pSyncMain$\,ns, $\sigma_\phi=\pPhiMain$\,deg, and
the biased blocked-LoS model.

\textit{Protocol.} Estimator parameters were tuned on seeds
\pSeedsTune{} and methods fixed on development seeds \pSeedsDev. An
earlier held-out set was consumed before the front/back disambiguation
was added, so all results come from a second, never-used set of ten
seeds (\pSeedsHeld), each simulated for two mounts and two traffic
densities. The seed is the statistical unit: we average its runs,
compare conditions with the exact two-sided Wilcoxon signed-rank test
over the ten seeds, and give \pCi\% bootstrap intervals over seeds.
The map-aided tracker was developed after the dominant-path estimator,
on the same tuning and development seeds, and was run once on the
held-out seeds with unchanged parameters.

\begin{figure}[t]
\centering
\includegraphics[width=\columnwidth]{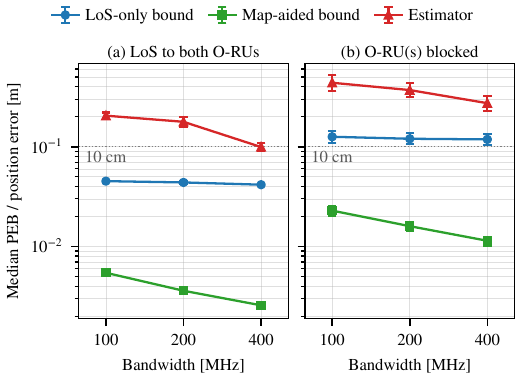}
\caption{Median single-epoch position error bound (LoS-only, map-aided)
and median dominant-path estimator error versus bandwidth, for epochs (a) with LoS
to both O-RUs and (b) with at least one blocked O-RU; main configuration (\pCi\% CI). The
bound and the error are different quantities (Section~IV). The LoS-only
bound barely changes with bandwidth because array calibration limits it
(Section~V-B).}
\label{fig:bw}
\end{figure}

\subsection{What Limits the Bound}
Fig.~\ref{fig:bw} shows the median bounds versus bandwidth. With LoS to
both O-RUs, the LoS-only PEB is below \pDecThr\,cm in
\pDecLosFourHundredLos{} of the epochs at \pBwC\,MHz and still in
\pDecLosHundredLos{} at \pBwA\,MHz. Relaxing one limit at a time at
\pBwB\,MHz scales the median LoS-only PEB (\pLimLosBase\,cm) by
\pLimLosBw{} for twice the bandwidth, \pLimLosSnr{} for ten times the
SNR, \pLimLosSync{} for perfect synchronization, and \pLimLosCal{} for
perfect calibration. On these short links, the delay and angle
measurements are precise enough that the element phase errors, which
bias the angles, dominate; the map-aided bound, in contrast, does improve with bandwidth (Fig.~\ref{fig:bw}). Without hardware errors, the LoS-only PEB is
below \pDecThr\,cm in \pHwLosIdeal{} of the epochs; with
\pSyncWorst\,ns and \pPhiWorst\,deg errors, in \pHwLosWorst. The
map-aided PEB is far less sensitive (\pHwMapIdeal{} and \pHwMapWorst),
because reflections from known surfaces add independent geometric
information.

\subsection{Blockage and the Digital Twin}
In the biased model, the median LoS-only PEB rises by a factor of
\pRatioLosB{} during blockage ($p=\pRatioPLosB$), and only
\pDecLosFourHundredBlk{} of the blocked epochs remain below \pDecThr\,cm.
If the attenuated LoS kept its geometry, the factor would be only
\pRatioLosA, which shows how optimistic it is to apply model~B's power
loss without its geometric consequence. Map-aided information lowers
the PEB during blockage to \pMapToLosBlk{} times the LoS-only value
($p=\pMapToLosP$). Along the street, the
LoS-only PEB varies with the geometry of the two O-RUs from about a
centimeter to over a decimeter, while the map-aided PEB stays mostly in
the millimeter range. A low map-aided PEB indicates that the information is present in the signal.

\subsection{Practical Estimators and the Bound}
At \pBwC\,MHz, the dominant-path estimator reaches a median error of
\pEstMedFourHundred\,cm (\pEstLosFourHundred\,cm with LoS to both O-RUs,
\pEstBlkFourHundred\,cm during blockage); \pEstDecFourHundred{} of its
errors are below \pDecThr\,cm, and its per-seed 90th percentile is
\pEstPninetyFourHundred\,m on average. At \pBwA\,MHz, the
median grows to \pEstMedHundred\,cm.

To compare the estimator with the bound on equal terms, we ran
\pValNumDraws{} independent noise and hardware draws through the
single-epoch estimator (without EKF) at \pValNumGeom{} fixed geometries
and decomposed its error as
\begin{equation}
\mathrm{RMSE}^2=\|\mathbb E[\hat{\mathbf p}]-\mathbf p\|^2
+\mathbb E\bigl[\|\hat{\mathbf p}-\mathbb E[\hat{\mathbf p}]\|^2\bigr],
\label{eq:decomp}
\end{equation}
i.e., squared bias plus squared centered RMS (Table~\ref{tab:val}). The
hardware errors are redrawn in every draw, so their contribution appears
in the centered RMS, whereas the bias contains only deterministic errors
such as unresolved multipath and the detoured LoS; within a run, a
hardware realization is fixed and acts as a bias. The squared bias accounts for most of the mean squared error in five of the
six geometries at \pBwC\,MHz (\pValBiasMseFour{} of it); in the
remaining one (lamppost, both LoS, $x=-20$\,m), the squared bias is
\pValBiasMseFourOther{} of it and the random spread is larger. The bias is caused by interference of unresolved multipath, and
in one geometry by a front/back mirror solution. The centered RMS is
geometry dependent (\pValCenToPebFour{} times the PEB at
\pBwC\,MHz): close to the bound at some geometries and an order of
magnitude above it at others. In blocked geometries, the centered RMS
falls below the PEB, which is possible because the estimator is biased: it uses the
detoured LoS precisely but in the wrong place. Over the held-out runs, front/back
mirror solutions remain in \pMirrorShareEpochs{} of the epochs and
dominate the squared error (RMS over all epochs \pMirrorARmse\,m), while reflected
dominant paths set the 90th percentile. A variant that used reflections
as virtual anchors did not reduce the tail on the development seeds
(90th percentile \pDevBPninety{} vs.\ \pDevAPninety\,m) and was not
pursued.

The map-aided tracker reaches a median error of \pTrkMed\,cm (\pTrkBlkMed\,cm
with at least one O-RU blocked) and a 90th percentile of
\pTrkPninety\,m, significantly better than the dominant-path estimator
(all $p=\pTrkVsAPAll$). Its largest errors
occur in the \SI{1}{s} after a UE re-enters the periodic street section,
a simulation artifact in which the track keeps the pre-jump position
until it re-initializes; these windows contain \pTrkWrapShare{} of the
epochs (not excluded).

\begin{table}[t]
\centering
\caption{Dominant-path estimator error around reference-cell blockage events
(10\,dB, main configuration).}
\label{tab:onset}
\setlength{\tabcolsep}{3pt}
\setlength{\aboverulesep}{0pt}\setlength{\belowrulesep}{0pt}
\setlength{\extrarowheight}{1.5pt}
\begin{tabular}{lcccc}
\toprule
\rowcolor{headblue}
Window & Median & 90th perc. & $>$\SI{0.1}{m} & $>$\SI{0.25}{m}\\
\midrule
1.0--0.5\,s before onset & \pOnsetPreOneMed\,cm & \pOnsetPreOnePninety\,m & \pOnsetPreOneShTen & \pOnsetPreOneShQuarter\\
0.5--0\,s before onset   & \pOnsetPreHalfMed\,cm & \pOnsetPreHalfPninety\,m & \pOnsetPreHalfShTen & \pOnsetPreHalfShQuarter\\
During the event         & \pOnsetDuringMed\,cm & \pOnsetDuringPninety\,m & \pOnsetDuringShTen & \pOnsetDuringShQuarter\\
Elsewhere                & \pOnsetElseMed\,cm & \pOnsetElsePninety\,m & \pOnsetElseShTen & \pOnsetElseShQuarter\\
\bottomrule
\end{tabular}
\end{table}

\begin{table}[t]
\centering
\caption{Planner outage minus A5 [s/UE-min] by source of the UE position,
with perfect blocker tracks. Negative values (shaded) mean less outage
than A5; $^\dagger$: not significant (seed-level Wilcoxon $p\ge0.05$;
pointwise tests, exploratory). Ref.\ = 3GPP reference budget
(\pRefMargin\,dB margin). $^\ast$Oracle diagnostic: mirror solutions
replaced by the single-epoch hypothesis closest to the true position.}
\label{tab:closing}
\newcommand{\cv}[1]{\expandafter\cvaux#1\relax}
\def\cvaux#1#2\relax{\if-#1\cellcolor{bettergreen}\ensuremath{-}\else\ensuremath{+}\fi#2}
\let\olddagger\dagger
\renewcommand{\dagger}{{\rlap{$\scriptstyle\olddagger$}}}
\setlength{\tabcolsep}{4pt}
\setlength{\aboverulesep}{0pt}\setlength{\belowrulesep}{0pt}
\setlength{\extrarowheight}{1.5pt}
\begin{tabular}{l*{5}{c}}
\toprule
\rowcolor{headblue}
 & \multicolumn{4}{c}{Link margin} & \\[-1pt]
\arrayrulecolor{black!40}\cmidrule(lr){2-5}\arrayrulecolor{black}
\rowcolor{headblue}
\multirow{-2}{*}{UE position} & 15\,dB & 20\,dB & 25\,dB & 30\,dB & \multirow{-2}{*}{Ref.}\\
\midrule
\rowcolor{groupgray}\multicolumn{6}{l}{\textit{Reference inputs}}\\
\quad exact               & \cv\pClosePerfFifteen & \cv\pClosePerfTwenty & \cv\pClosePerfTwentyFive & \cv\pClosePerfThirty & \cv\pClosePerfRef\\
\quad white \SI{1}{m} error & \cv\pCloseWhiteFifteen & \cv\pCloseWhiteTwenty & \cv\pCloseWhiteTwentyFive & \cv\pCloseWhiteThirty & \cv\pCloseWhiteRef\\
\rowcolor{groupgray}\multicolumn{6}{l}{\textit{Synthetic, bound level (full covariance)}}\\
\quad map-aided bound     & \cv\pCloseFullMapFifteen & \cv\pCloseFullMapTwenty & \cv\pCloseFullMapTwentyFive & \cv\pCloseFullMapThirty & \cv\pCloseFullMapRef\\
\quad LoS-only bound      & \cv\pCloseFullLosFifteen & \cv\pCloseFullLosTwenty & \cv\pCloseFullLosTwentyFive & \cv\pCloseFullLosThirty & \cv\pCloseFullLosRef\\
\quad LoS-only, no fix held & \cv\pCloseFullLosHeldFifteen & \cv\pCloseFullLosHeldTwenty & \cv\pCloseFullLosHeldTwentyFive & \cv\pCloseFullLosHeldThirty & \cv\pCloseFullLosHeldRef\\
\rowcolor{groupgray}\multicolumn{6}{l}{\textit{Estimated}}\\
\quad map-aided tracker   & \cv\pCloseTrkFifteen & \cv\pCloseTrkTwenty & \cv\pCloseTrkTwentyFive & \cv\pCloseTrkThirty & \cv\pCloseTrkRef\\
\quad dominant-path       & \cv\pCloseEstFifteen & \cv\pCloseEstTwenty & \cv\pCloseEstTwentyFive & \cv\pCloseEstThirty & \cv\pCloseEstRef\\
\quad ideal hardware      & \cv\pCloseEstIdealFifteen & \cv\pCloseEstIdealTwenty & \cv\pCloseEstIdealTwentyFive & \cv\pCloseEstIdealThirty & \cv\pCloseEstIdealRef\\
\quad capped at p90       & \cv\pCloseEstCapFifteen & \cv\pCloseEstCapTwenty & \cv\pCloseEstCapTwentyFive & \cv\pCloseEstCapThirty & \cv\pCloseEstCapRef\\
\quad mirror-corr.$^\ast$ & \cv\pCloseOracleFifteen & \cv\pCloseOracleTwenty & \cv\pCloseOracleTwentyFive & \cv\pCloseOracleThirty & \cv\pCloseOracleRef\\
\bottomrule
\end{tabular}
\end{table}

\subsection{Positioning Errors Around Blockage Onsets}
Table~\ref{tab:onset} shows the estimator error around the
\pOnsetNumEvents{} blockage events of the reference cell, defined per UE
as the cell with the larger unblocked received power; an event is an
interval with LoS loss of at least \SI{10}{dB}, with gaps below
\SI{0.5}{s} merged; each epoch is counted in one window, with priority
to the event, and the first \SI{2}{s} of each run are excluded. In the second before onset, a window relevant to proactive
handover, the median error and the share of errors above
\SI{0.25}{m} are about as elsewhere, but the large errors become larger:
the 90th percentile rises from \pOnsetElsePninety{} to
\pOnsetPreHalfPninety\,m. During the event, most of the distribution
deteriorates, with the median growing to \pOnsetDuringMed\,cm. The overall median hides both effects. For the map-aided tracker, the
90th percentile does not rise before onset (\pTrkOnsetPreHalfPninety{}
vs.\ \pTrkOnsetElsePninety\,m elsewhere), and its median during the
event is \pTrkOnsetDuringMed\,cm. The windows refer to the blockage
events of a reference cell, not to the decision times of a particular
planner.

\subsection{Implications for Handover}
Table~\ref{tab:closing} feeds positions into the planner, with perfect blocker tracks so that the user position is
the only source of error. With exact positions, the planner has a lower
outage than A5 from \pMarginFifteen\,dB margin upward (significantly at
15 and \SI{25}{dB} and at the reference). As a sensitivity test, we feed
synthetic errors $\mathbf e_t\sim\mathcal N\bigl(\mathbf 0,\mathbf
J_{\mathbf p}(t)^{-1}\bigr)$, drawn independently per epoch from the
single-epoch bound of the respective model, which is anisotropic
(median ratio of principal standard deviations \pAniso). Errors at the
map-aided bound preserve a lower mean outage than A5 at most tested
margins, significantly at \SI{15}{dB} (at \SI{25}{dB} only without correction); errors at
the LoS-only bound, which grow during blockage, make it worse than A5 at
every margin. With the positions of the dominant-path estimator, the
planner is significantly worse than A5 at every margin, also with ideal
hardware; with those of the map-aided tracker, it is not significantly
different from A5 at any margin from \SI{15}{dB} up ($p\ge\pTrkMinP$),
although at \SI{25}{dB} its bootstrap interval lies above zero; at
\SI{10}{dB} it is still significantly worse (\pCloseTrkTen~s/UE-min,
$p=\pCloseTrkTenP$), as at all lower margins ($p\le\pTrkLowMaxP$). Capping each error of the EKF output at the
pooled 90th percentile (\pCapPninety\,m) or replacing its mirror
solutions by the single-epoch hypothesis closest to the true position
(an oracle diagnostic, not refiltered) reduces the loss but leaves it
significant, so the largest errors are not its only
cause. With Holm correction across the margins of each row, all these
significant results remain significant except the map-aided bound at
\SI{25}{dB} (adjusted $p=\pHolmMapTwentyFiveP$).

Synthetic bound-level errors are an optimistic model; they show what
positioning would have to achieve, not that a practical estimator
achieves it. In the \pDualBlockShare{} of epochs in which both O-RUs are
blocked, the LoS-only bound is singular; these epochs are treated as
having no fix (\SI{100}{m} standard deviation). Holding the last valid
position instead keeps the LoS-only row above A5 at every margin but
removes its significance, so that row depends mainly on how missing
fixes are handled. The synthetic errors are independent
across epochs, whereas hardware errors persist over a run; the effect of
temporal correlation on the planner is not evaluated here. The absolute gains at stake are small: at \SI{15}{dB} margin, A5 has an
outage of \pAfiveFifteen~s/UE-min, and the relative change is
\mbox{\pRelPerfFifteen} with exact positions and \mbox{\pRelMapFifteen} with
bound-level map-aided positions; at the reference, A5 has
\pAfiveRef~s/UE-min, and the changes are \mbox{\pRelPerfRef} and
\mbox{\pRelMapRef}, respectively; the application-level impact is not evaluated here.

\section{Discussion and Limitations}
For blockage-aware handover, what matters is the error around blockage
onsets, not the overall median. The dominant-path estimator falls short
of the bound mainly because of systematic multipath errors, which a
simple virtual-anchor variant did not remove. The map-aided tracker
resolves several paths and associates them probabilistically with known
surfaces; it cuts the median error about threefold but remains well
above the map-aided bound. With its positions, proactive handover is no
longer significantly worse than A5 from \SI{15}{dB} up, yet it does not
reach the small gain of exact positions. Whether this gap comes from the
tracker's residual errors, including those after the UE jumps, remains
open.

The study has limitations. It considers one street canyon with two
O-RUs, specular paths of order at most \pMaxDepth{} without diffuse
scattering, and fully digital arrays read out in one symbol; their
isotropic elements cause a front/back ambiguity that a back plane would
remove \cite{dong2026deployable}. The map is exact in both the
map-aided bound and the tracker, which was added after the first
evaluation. A blocked LoS follows our biased model, and bodies are
opaque knife-edge screens, without window transmission, vehicle-body
reflections, or self-blockage by the user. Finally, the bounds are
single-epoch and ignore the temporal information of tracking.

\section{Conclusion}
We asked whether the radio units that serve a user can localize it
accurately enough for blockage-aware handover, in a short-range street
canyon with two O-RUs. With LoS to both O-RUs, array calibration rather
than bandwidth limits the LoS-only bound. Blockage degrades this bound
substantially, whereas known reflected paths keep the map-aided bound
much lower. A dominant-path estimator makes mostly systematic multipath
errors that worsen around blockage, and its positions make proactive
handover worse than reactive handover. A map-aided multipath tracker
cuts the median error to a few centimeters and keeps it low around
blockage onsets; with its positions, proactive handover is no longer
significantly worse than the 3GPP event from \SI{15}{dB} up. Reaching
the small gain that exact positions provide remains open.

\section*{Acknowledgment}
This material is based upon work supported in part by NSF under Awards CNS-2120442 and IIS-2325863, and NTIA under Award No. 51-60-IF007. Any opinions, findings, and conclusions or recommendations expressed in this publication are those of the author(s) and do not necessarily reflect the views of the NSF and NTIA. AI assisted with coding and enhancing the text; the authors verified all content.

\bibliographystyle{IEEEtran}
\bibliography{refs2}

\end{document}

%% file: numbers2.tex
\newcommand{\pAfiveFifteen}{1.88}
\newcommand{\pAfiveRef}{0.058}
\newcommand{\pAniso}{3.0}
\newcommand{\pArray}{\ensuremath{8\times8}}

\newcommand{\pBlockedDb}{10}

\newcommand{\pBreakEven}{8--13}

\newcommand{\pBreakEvenRms}{11--18}

\newcommand{\pBwA}{100}
\newcommand{\pBwB}{200}
\newcommand{\pBwC}{400}
\newcommand{\pCapPninety}{1.90}

\newcommand{\pCarrier}{28}
\newcommand{\pCi}{95}

\newcommand{\pCloseEstCapFifteen}{+0.525}
\newcommand{\pCloseEstCapRef}{+0.051}
\newcommand{\pCloseEstCapThirty}{+0.118}
\newcommand{\pCloseEstCapTwenty}{+0.434}
\newcommand{\pCloseEstCapTwentyFive}{+0.251}
\newcommand{\pCloseEstFifteen}{+0.652}
\newcommand{\pCloseEstIdealFifteen}{+0.584}
\newcommand{\pCloseEstIdealRef}{+0.054}
\newcommand{\pCloseEstIdealThirty}{+0.121}
\newcommand{\pCloseEstIdealTwenty}{+0.483}
\newcommand{\pCloseEstIdealTwentyFive}{+0.295}
\newcommand{\pCloseEstRef}{+0.052}
\newcommand{\pCloseEstThirty}{+0.118}
\newcommand{\pCloseEstTwenty}{+0.507}
\newcommand{\pCloseEstTwentyFive}{+0.284}
\newcommand{\pCloseFullLosFifteen}{+0.162}
\newcommand{\pCloseFullLosHeldFifteen}{+0.071$^\dagger$}
\newcommand{\pCloseFullLosHeldRef}{+0.036$^\dagger$}
\newcommand{\pCloseFullLosHeldThirty}{+0.073$^\dagger$}
\newcommand{\pCloseFullLosHeldTwenty}{+0.142$^\dagger$}
\newcommand{\pCloseFullLosHeldTwentyFive}{+0.058$^\dagger$}
\newcommand{\pCloseFullLosRef}{+0.097}
\newcommand{\pCloseFullLosThirty}{+0.151}
\newcommand{\pCloseFullLosTwenty}{+0.301}
\newcommand{\pCloseFullLosTwentyFive}{+0.175}
\newcommand{\pCloseFullMapFifteen}{-0.099}
\newcommand{\pCloseFullMapRef}{-0.005$^\dagger$}
\newcommand{\pCloseFullMapThirty}{+0.021$^\dagger$}
\newcommand{\pCloseFullMapTwenty}{-0.049$^\dagger$}
\newcommand{\pCloseFullMapTwentyFive}{-0.044}
\newcommand{\pCloseOracleFifteen}{+0.461}
\newcommand{\pCloseOracleRef}{+0.048}
\newcommand{\pCloseOracleThirty}{+0.115}
\newcommand{\pCloseOracleTwenty}{+0.443}
\newcommand{\pCloseOracleTwentyFive}{+0.260}

\newcommand{\pClosePerfFifteen}{-0.108}
\newcommand{\pClosePerfRef}{-0.025}
\newcommand{\pClosePerfThirty}{-0.007$^\dagger$}
\newcommand{\pClosePerfTwenty}{-0.066$^\dagger$}
\newcommand{\pClosePerfTwentyFive}{-0.061}
\newcommand{\pCloseTrkFifteen}{+0.009$^\dagger$}
\newcommand{\pCloseTrkRef}{-0.002$^\dagger$}
\newcommand{\pCloseTrkTen}{+0.414}
\newcommand{\pCloseTrkTenP}{0.0059}
\newcommand{\pCloseTrkThirty}{+0.015$^\dagger$}
\newcommand{\pCloseTrkTwenty}{+0.031$^\dagger$}
\newcommand{\pCloseTrkTwentyFive}{+0.072$^\dagger$}
\newcommand{\pCloseWhiteFifteen}{+1.402}
\newcommand{\pCloseWhiteRef}{+0.117}
\newcommand{\pCloseWhiteThirty}{+0.290}
\newcommand{\pCloseWhiteTwenty}{+0.824}
\newcommand{\pCloseWhiteTwentyFive}{+0.476}

\newcommand{\pDecLosFourHundredBlk}{47\%}
\newcommand{\pDecLosFourHundredLos}{84\%}

\newcommand{\pDecLosHundredLos}{81\%}

\newcommand{\pDecThr}{10}

\newcommand{\pDerivFdMaxRel}{\ensuremath{3.2\times 10^{-6}}}
\newcommand{\pDerivFdNumUe}{1000}

\newcommand{\pDevAPninety}{1.87}

\newcommand{\pDevBPninety}{3.52}

\newcommand{\pDualBlockShare}{3.0\%}
\newcommand{\pEpoch}{0.1}

\newcommand{\pEstBlkFourHundred}{27}

\newcommand{\pEstDecFourHundred}{45\%}

\newcommand{\pEstLosFourHundred}{10}

\newcommand{\pEstMedFourHundred}{12}
\newcommand{\pEstMedHundred}{24}

\newcommand{\pEstPninetyFourHundred}{2.10}

\newcommand{\pFimJobErr}{\ensuremath{2.9\times 10^{-6}}}
\newcommand{\pFimToyErr}{\ensuremath{6.6\times 10^{-10}}}

\newcommand{\pHolmMapTwentyFiveP}{0.055}

\newcommand{\pHwLosIdeal}{92\%}

\newcommand{\pHwLosWorst}{44\%}

\newcommand{\pHwMapIdeal}{99\%}

\newcommand{\pHwMapWorst}{98\%}
\newcommand{\pLaneWidth}{3.5}

\newcommand{\pLimLosBase}{4.8}
\newcommand{\pLimLosBw}{0.97}
\newcommand{\pLimLosCal}{0.07}
\newcommand{\pLimLosSnr}{0.94}
\newcommand{\pLimLosSync}{0.72}

\newcommand{\pLoopDelay}{20}
\newcommand{\pMapToLosBlk}{0.09}
\newcommand{\pMapToLosP}{0.002}
\newcommand{\pMarginFifteen}{15}

\newcommand{\pMaxDepth}{2}
\newcommand{\pMaxPaths}{8}

\newcommand{\pMirrorARmse}{6.36}

\newcommand{\pMirrorShareEpochs}{5\%}
\newcommand{\pNoiseFig}{7}
\newcommand{\pNscA}{792}
\newcommand{\pNscB}{1584}
\newcommand{\pNscC}{3168}

\newcommand{\pNumerology}{3}
\newcommand{\pOnsetDuringMed}{87}
\newcommand{\pOnsetDuringPninety}{3.26}
\newcommand{\pOnsetDuringShQuarter}{72\%}
\newcommand{\pOnsetDuringShTen}{88\%}
\newcommand{\pOnsetElseMed}{10}
\newcommand{\pOnsetElsePninety}{1.59}
\newcommand{\pOnsetElseShQuarter}{29\%}
\newcommand{\pOnsetElseShTen}{49\%}
\newcommand{\pOnsetNumEvents}{752}
\newcommand{\pOnsetPreHalfMed}{11}
\newcommand{\pOnsetPreHalfPninety}{4.49}
\newcommand{\pOnsetPreHalfShQuarter}{29\%}
\newcommand{\pOnsetPreHalfShTen}{53\%}
\newcommand{\pOnsetPreOneMed}{12}
\newcommand{\pOnsetPreOnePninety}{4.24}
\newcommand{\pOnsetPreOneShQuarter}{30\%}
\newcommand{\pOnsetPreOneShTen}{54\%}
\newcommand{\pOruOffset}{40}

\newcommand{\pOruOneLampZ}{5}

\newcommand{\pOruZeroFacZ}{8}

\newcommand{\pOruZeroLampZ}{5}
\newcommand{\pOverhead}{2.3\%}

\newcommand{\pPedCrossingHigh}{4}
\newcommand{\pPedCrossingLow}{2}

\newcommand{\pPedWalkersHigh}{12}
\newcommand{\pPedWalkersLow}{6}
\newcommand{\pPhiList}{0, 2 and 5}
\newcommand{\pPhiMain}{2}
\newcommand{\pPhiWorst}{5}
\newcommand{\pPlanH}{0.5--3}
\newcommand{\pRatioLosA}{1.2}
\newcommand{\pRatioLosB}{2.8}

\newcommand{\pRatioPLosB}{0.002}

\newcommand{\pRefMargin}{35.9}
\newcommand{\pRelMapFifteen}{\ensuremath{-}5.2\%}
\newcommand{\pRelMapRef}{\ensuremath{-}8.2\%$^\dagger$}
\newcommand{\pRelPerfFifteen}{\ensuremath{-}5.7\%}
\newcommand{\pRelPerfRef}{\ensuremath{-}43.1\%}
\newcommand{\pRunDuration}{60}
\newcommand{\pSNRreq}{9.3}
\newcommand{\pSameCellLoss}{26--33}
\newcommand{\pScs}{120}
\newcommand{\pSeedsDev}{1001--1010}
\newcommand{\pSeedsHeld}{3001--3010}
\newcommand{\pSeedsTune}{101--105}

\newcommand{\pStreetPeriod}{80}
\newcommand{\pStreetWidth}{18.2}

\newcommand{\pSyncListShown}{0, 1 and 3}
\newcommand{\pSyncMain}{1}
\newcommand{\pSyncWorst}{3}
\newcommand{\pTauHO}{20}
\newcommand{\pTrkBlkMed}{4.5}

\newcommand{\pTrkLowMaxP}{0.0059}
\newcommand{\pTrkMed}{3.7}
\newcommand{\pTrkMinP}{0.084}
\newcommand{\pTrkOnsetDuringMed}{6.4}

\newcommand{\pTrkOnsetElsePninety}{0.35}

\newcommand{\pTrkOnsetPreHalfPninety}{0.20}

\newcommand{\pTrkPninety}{0.25}
\newcommand{\pTrkVsAPAll}{0.002}

\newcommand{\pTrkWrapShare}{1.7\%}

\newcommand{\pTxPower}{23}
\newcommand{\pUeHeight}{1.5}
\newcommand{\pUeSpeed}{1.0--1.2}

\newcommand{\pValBiasMseFour}{98--100\%}
\newcommand{\pValBiasMseFourOther}{38\%}

\newcommand{\pValCenToPebFour}{0.077--13}

\newcommand{\pValNumDraws}{500}
\newcommand{\pValNumGeom}{6}

\newcommand{\pVehLaneKmHigh}{100}
\newcommand{\pVehLaneKmLow}{37.5}

\newcommand{\pVehSpeed}{6--12}